\documentclass{article}
\date{}
\usepackage{amssymb,amsmath,amsthm}

\usepackage{latexsym} 
\usepackage{alltt}
\usepackage[all]{xy} 
\CompileMatrices 
\xyoption{frame}

\usepackage{amscd} 
\usepackage{latexsym}
\usepackage{epsf}
\usepackage{float}
\floatstyle{ruled}
\newfloat{Data Module}{thp}{lop}[section]
\usepackage{verbatim}
\usepackage{moreverb}
\usepackage{alltt}
\usepackage{amsmath}
\usepackage{graphics}
\usepackage{amssymb}
\usepackage{bbold}
\theoremstyle{plain}

\theoremstyle{definition}

\begin{document}
\title{Note on paraconsistency and reasoning about fractions}
\author{Jan A. Bergstra and Inge Bethke$^{\ast}$\thanks{$^\ast$Corresponding author. Email:I.Bethke@uva.nl}\\
Informatics Institute, University of Amsterdam\\
Science Park 904, 1098 XH Amsterdam,\\
 The Netherlands
}
\maketitle
\begin{abstract}
We apply a paraconsistent  strategy to reason about fractions.
\end{abstract}

\section{Introduction} 
Suppose we want to define an arithmetic framework in which it is possible to reason about fractions in a consistent and reliable way and in which the usual laws of arithmetic hold:
e.g.\ for natural numbers $n,l$ and $k \neq 0\neq m$ the equations
\[
\frac{n}{m} + \frac{l}{k}= \frac{nk +lm}{mk} \ \ \ \ \ \text{ and }\ \ \ \ \ 
\frac{n}{m} +\frac{l}{m}=\frac{n+l}{m}
\]
should be valid. At the same time, we want to consider fractions as mathematical expressions with typical syntactic operations like the numerator $num(\ )$ satisfying 
\[
num(\frac{n}{m})=n.
\]
Then equational logic dictates 
\[
n+l= num(\frac{n+l}{m})=num(\frac{n}{m} +\frac{l}{m})=num( \frac{nm +lm}{mm})=nm +lm.
\]
for arbitrary $l, n$ and  $m\neq 0$, and our framework is inconsistent. 
Nevertheless, fractions are of great practical and abstract importance and we should be able to reason about them without lapsing into absurdity.

In mathematics education, one can get around this predicament in various ways: avoiding the concepts of numerator and denominator (see e.g.\  \cite{Rollnik}), viewing fractions as heterogeneous subject (see e.g.\  \cite{Padberg}), or accepting cognitive conflicts (see e.g.\ \cite{Tall}). In this note, we propose to apply paraconsistent reasoning. 

A \emph{paraconsistent} logic is a way to reason about inconsistent information without \emph{exploding} in the sense that if a contradiction is obtained, then everything can be obtained. Paraconsistent logics come in a broad spectrum, ranging from logics with the thought that if a contradiction were true, then everything would be true, to logics that claim that some contradictions really are true. In this note, we choose a particular paraconsistent reasoning strategy to tackle the dilemma sketched above. We do not claim that this is the only possible way how to proceed in our  scenario; other paraconsistent approaches may be suitable as well.

\section{The C \& P structure}
The approach taken here belongs to the \emph{preservationist} school. The fundamental idea is that, given an inconsistent collection of premises, one should not try to reason about the collection of premises as a whole, but rather focus on internally consistent subsets of premises. In 2004, Brown and Priest \cite{BP04} introduced the \emph{Chunk and Permeate} (C \& P) strategy for dealing with reasoning situations involving incompatible assumptions. In this reasoning strategy, a theory is broken up into chunks and only restricted information is allowed to pass from one chunk to another. 
In what follows we give a precise rendering of this idea.

Let $L$ be some classical language and let $\vdash$ be an appropriate classical consequence relation. If $\Gamma$ is a set of sentences in $L$, $\Gamma^{\vdash}$ denotes  the closure of $\Gamma$ under $\vdash$. A \emph{covering} of $\Gamma$ is a set $\{\Gamma_i\mid i\in I\}$ of classical consistent sets of sentences such that  $\Gamma=\bigcup_{i\in I} \Gamma_i$. If $C=\{\Gamma_i\mid i\in I\}$ is a covering of $\Gamma$,  a \emph{permeability relation} $\rho$ on $C$ is a map from $I \times I$ to sets of sentences of $L$. If $i_0\in I$, $\langle C, \rho, i_0\rangle$ is called a \emph{C \& P structure} on $\Gamma$.

If $\mathcal{P}=\langle C, \rho, i_0\rangle$ is a C \& P structure on $\Gamma$ and $A$ is a sentence in the language under consideration,  then
\[
\Gamma \vdash_{\mathcal{P}} A \text{ if and only if } A\in \Gamma^\omega_{i_0}
\]
where 
\[
\begin{array}{rcl}
\Gamma_i^0&=& \Gamma^{\vdash}_i\\
\Gamma_i^{n+1} &=& (\Gamma^n_i \cup \bigcup_{j\in I}(\Gamma_j^n \cap \rho(j,i)))^{\vdash}.
\end{array}
\]
The C \& P consequences of $\Gamma$ are thus sentences that can be inferred in the designated chunk $i_0$ when information has been allowed to flow from other chunks along the permeability relation. 

In our case, a simple binary C \& P structure will be sufficient. We partition our inconsistent theory into two chunks: the \emph{source} chunk $\Gamma_S$ and the \emph{target} chunk $\Gamma_T$. The flow of information is from the former to the latter only and the target chunk is the output chunk. Note that in this case, $\Gamma^\omega_S=\Gamma^\vdash_S$ and hence 
\[
\Gamma^\omega_T=\Gamma^1_T= (\Gamma_T \cup (\Gamma^\vdash_S\cap \rho(S,T)))^\vdash.
\]

We let $L$ be the language of equational logic
with terms over the two-sorted signature  $\Sigma$ containing $0, 1$,  the numerator and denominator $num$ and $denom$, the fraction $\frac{\  \ }{\  \ }$, symbols for addition and multiplication of natural numbers and fractions, respectively, and an additional error element $a$ produced by division by zero. The inconsistent theory $\Gamma$ is the equational theory of the set of axioms given in Table \ref{gamma}. Here $k,l,m,n$ range over natural numbers, and $\alpha, \beta, \gamma$ denote fractions. 
\begin{table}
\[
\begin{array}{lrclll}
\hline
\\
&n+0&=&n&&(1)\\[1 mm]
&(n+m)+l&=&n+(m+l)&&(2)\\[1 mm]
&n+m&=&m+n&&(3)\\[1 mm]
&n\cdot 1&=&n&&(4)\\[1 mm]
&(n\cdot m)\cdot l&=&n\cdot (m\cdot l)&&(5)\\[1 mm]
&n\cdot m&=&m\cdot n&&(6)\\[1 mm]
&n\cdot (m+l)&=&n\cdot m +n \cdot l&&(7)\\[1 mm]
m\neq 0\neq k \rightarrow  &\dfrac{n}{m}\cdot \dfrac{l}{k}&=&\dfrac{n\cdot l}{m\cdot k}&&(8)\\[3 mm]
m\neq 0\neq k \rightarrow &\dfrac{n}{m} + \dfrac{l}{k}&=&\dfrac{n\cdot k+l\cdot m}{m \cdot k}&&(9)\\[4 mm]
m\neq 0\rightarrow &\dfrac{n}{m} + \dfrac{l}{m}&=&\dfrac{n+l}{m}&&(10)\\[4 mm]
&(\alpha +\beta)+\gamma&=&\alpha+(\beta +\gamma)&&(11)\\[1 mm]
&\alpha+\beta&=&\beta +\alpha&&(12)\\[1 mm]
&(\alpha  \cdot \beta)\cdot \gamma&=&\alpha\cdot (\beta \cdot \gamma)&&(13)\\[1 mm]
&\alpha\cdot \beta&=&\beta \cdot \alpha&&(14)\\[1 mm]
&\alpha \cdot (\beta +\gamma )&=&\alpha\cdot \beta +\alpha \cdot \gamma&&(15)\\[2 mm]
m\neq 0\ \wedge\ m\neq a\rightarrow &num(\dfrac{n}{m})&=&n&&(16)\\[3 mm]
m\neq 0\  \wedge\ n\neq a\rightarrow &denom(\dfrac{n}{m})&=&m&&(17)\\
\\
\hline
\end{array}
\]
\caption{The inconsistent equational theory $\Gamma$}\label{gamma}
\end{table}
The binary C \& P structure for our problem is $\langle \{\Gamma_S,\Gamma_T\}, \rho, T\rangle$ where
\begin{itemize}
\item the source chunk $\Gamma_S$ is the theory of the axioms (1) -- (8), (11) -- (17) of Table \ref{gamma} and the axioms (18) and (19) of Table \ref{alternative}, 
\item the target chunk $\Gamma_T$ consists of the axiom (20) of Table  \ref{alternative}, and
\item the information $\rho$ that is allowed to flow from source to target consists of the axioms of the source except for (16) and (17).
\end{itemize}

\begin{table}
\[
\begin{array}{lrclll}
\hline
\\
m\neq 0 \rightarrow &\dfrac{n}{m} &=&\dfrac{n}{1} \cdot \dfrac{1}{m}&&(18)\\[4 mm]
&\dfrac{n}{1} + \dfrac{m}{1}&=&\dfrac{n+m}{1}&&(19)\\[4 mm]
m\neq 0\neq k \rightarrow &\dfrac{n\cdot k}{m\cdot k}&=&\dfrac{n}{m}&&(20)\\
\\
\hline
\end{array}
\]
\caption{Alternative axioms for source and target}\label{alternative}
\end{table}

Observe that in the source, fractions with identical denominators can be added $(\dag)$: for $m\neq 0$ we have
\[
\begin{array}{rcll}
\dfrac{n}{m}+\dfrac{k}{m} &=&\dfrac{n}{1}\cdot \dfrac{1}{m} + \dfrac{k}{1}\cdot \dfrac{1}{m}& (18)\\[3 mm]
&=&(\dfrac{n}{1} + \dfrac{k}{1}) \cdot \dfrac{1}{m}&(6), (7)\\[3 mm]
&=&\dfrac{n+k}{1}\cdot \dfrac{1}{m}&(19)\\[3 mm]
&=&\dfrac{n+k}{m}&(18)
\end{array}
\]
This fact together with axiom (20) yields axiom (9). Thus $\{\Gamma_S, \Gamma_T\}$ covers $\Gamma$.

That $\Gamma_S$ is consistent can be seen as follows. We let $\mathbb{M}$ be the $\Sigma$-algebra with the sorts $\mathbb{N}_a =\mathbb{N}\cup \{a\}$ and $\mathbb{F}=\mathbb{N}\times \mathbb{N}$ where the operations are interpreted as follows. 
If $\circ \in \{+,\cdot\}$, then $\circ:\mathbb{N}_a \times \mathbb{N}_a \rightarrow \mathbb{N}_a$ is defined by
\[
x\circ y=
\begin{cases}
x\circ y &\text{ if $x,y\in \mathbb{N}$}\\
a &\text{ otherwise},
\end{cases}
\]
and $\dfrac{\  \ }{\  \ }:\mathbb{N}_a \times \mathbb{N}_a \rightarrow \mathbb{F}$ is defined by
 \[
\dfrac{x }{y }=
\begin{cases}
(x, y) &\text{ if $x\in \mathbb{N}, y\in \mathbb{N}$}\\
(0,0) &\text{ otherwise},
\end{cases}
\]
$num, denom: \mathbb{F} \rightarrow \mathbb{N}_a$  are defined by 
\[
num((n,m))=
\begin{cases}
n&\text{ if $m\neq 0$}\\
a &\text{ otherwise},
\end{cases}
\]
and 
\[
denom((n,m))=
\begin{cases}
m &\text{ if $m\neq 0$}\\
a &\text{ otherwise},
\end{cases}
\]
and $+,\cdot :\mathbb{F} \times \mathbb{F} \rightarrow \mathbb{F}$  are defined by 
 \[
(n,m)+(l,k)=
\begin{cases}
(n+l,m) &\text{ if $m=k\neq 0$}\\
(0,0) &\text{ otherwise}, 
\end{cases}
\]
and 
 \[
(n,m)\cdot (l,k)=
\begin{cases}
(nl,mk) &\text{ if $m\neq 0\neq k$}\\
(0,0) &\text{ otherwise}.
\end{cases}
\]
Note that $\mathbb{M}$ does not satisfy the full axiom of fraction addition (9). It is easy to see that $\mathbb{M}$ is a model for $\Gamma_S$. Clearly, $\Gamma_T$ is consistent. 
$\Gamma_T\cup \rho$ is consistent too: $\mathbb{Q}_0$ and $\mathbb{Q}_a$ |the zero-totalized and the $a$-totalized meadow of the rational numbers  (see e.g.\ \cite{BBP2013, BP2014}), respectively|are both models of $\Gamma_T\cup \rho$. 
We have thus arrived at a consistent theory where
we have full addition of fractions: for $m\neq 0\neq l$
\[
\begin{array}{rcll}
\dfrac{n}{m} + \dfrac{k}{l}&=& \dfrac{nl}{ml} + \dfrac{km}{lm}& (20)\\[3 mm]
&=&\dfrac{nl}{ml} + \dfrac{km}{ml}&(6)\\[3 mm]
&=&\dfrac{nl + km}{ml}&(\dag).
\end{array}
\]

\section{Conclusion}
$\mathbb{M}$ is only one of a number of possible structures that incorporate fractions in such a manner that the oprations numerator and denominator can be properly defined. We do not exclude that more convincing datatypes can be found as models for $\Gamma_S$.
For instance with $/$ representing integer division, the defining equation 
\[
\dfrac{n}{m} + \dfrac{k}{l} = \dfrac{(n \cdot l + k \cdot m)/ gcd(m,l)}{m \cdot l/ gcd(m,l)}
\]
is compatible with having functions $num$ and $denom$ around in combination with associativity and the rule
\[
\dfrac{n}{m} + \dfrac{k}{m} = \dfrac{n+k}{m}.
\]
We will not explore such alternatives in this note. For meadows different from common meadows see e.g. \cite{BM2015}.

The relation between the views $\Gamma_S$ and $\Gamma_T$ requires further attention. Following the original line of thought,  $\Gamma_S$ is a stepping stone towards $\Gamma_T$ which in our case implies that the application of the C \& P strategy does not lead  to a picture where fractions and rationals coexist with the operations $num$ and $denom$.

It seems plausible to imagine that a person thinking along these lines about fractions switches back and forth between $\Gamma_S$ and $\Gamma_T$. The situation may then be compared with how a human observer looks at the Necker cube \cite{Mo06,N1832}. This cube is a 2-dimensional figure which invites an observer to ``see" in an alternating manner two different 3-dimensional interpretations of it. Instead of the 3rd dimension we consider a logic, informally understood. And $\Gamma_S$ and $\Gamma_T$ represent two logics of fractions the alternation of which seems to be adequately described by the inconsistency-tolerant technique of C \& P.


\begin{thebibliography}{99}

\bibitem{BBP2013}
Bergstra, J.A.,  Bethke, I., and Ponse, A. (2013).
\newblock Cancellation meadows: a generic basis theorem and some applications.
\newblock {\em The Computer Journal}, 56(1):3--14, \texttt{doi:10.1093/comjnl/bsx147}. 
(Also  available at arXiv:0803.3969 [math.RA, cs.LO].)

\bibitem{BM2015}
Bergstra, J.A. and Middelburg, C.A. (2015).
\newblock Division by zero in non-involutive meadows .
\newblock {\em Journal of Applied Logic}, 13(1):1--12, \texttt{doi:10.1016/j.jal.2014.10.001}. 
(Also  available at arXiv:1406.2092 [math.RA, cs.LO].)

\bibitem{BP2014}
Bergstra, J.A. and Ponse, A. (2015).
\newblock Division by zero in common meadows.
\newblock In de Nicola, R. and Hennicker, R. (eds.), \emph{Software Services and Systems}, LNCS 8950, pp. 46--61. Springer.
(Also  available at arXiv:1406.3280v2 [cs.LO].)


\bibitem{BP04}
Brown, B. and Priest, G. (2004).
\newblock Chunk and Permeate, a paraconsistent inference strategy. Part I: the infinitesimal calculus.
\newblock {\em Journal of Philosophical Logic}, 33:379 --388.

%\bibitem{M11}
%Middelburg, C.A. (2011).
%\newblock \emph{A Survey of Paraconsistent Logics}. 
%\newblock arXiv:1103.4324[cs.LO].

\bibitem{Mo06}
Mortensen, Ch.  (2006).
\newblock An Analysis of Inconsistent and Incomplete Necker Cubes.
\newblock \emph{Australasian Journal of Logic}, 4:216--225.



\bibitem{N1832}
Necker, L.A. (1832).
\newblock Observations on some remarkable optical phaenomena seen in Switzerland; and on an optical phaenomenon which occurs on viewing a figure of a crystal or geometrical solid.
\newblock \emph{London and Edinburgh Philosophical Magazine and Journal of Science}, 1(5):329--337.

\bibitem{Padberg}
Padberg, F. (2012).
\newblock \emph{Didaktik der Bruchrechnung} ($4_{th}$ edition).
\newblock Series: Mathematik, Primar- und Sekundarstufe, Springer-Spektrum.

\bibitem{Rollnik}
Rollnik, S. (2009).
\newblock \emph {Das pragmatische Konzept f\"ur den Bruchrechenunterricht}.
\newblock PhD thesis, University of Flensburg, Germany.

\bibitem{Tall}
Tall, D. and Vinner, S. (1981).
\newblock Concept image and concept definition in mathematics, with special reference to limits and continuity.
\newblock {\em Educational Studies in Mathematics}, 12 151--169.

\end{thebibliography}
\end{document}